# Resonant electron attachment to mixed hydrogen/oxygen and deuterium/oxygen clusters

Michael Renzler,[1,2] Lorenz Kranabetter,[1] Erik Barwa,[1] Lukas Grubwieser,[1] Paul Scheier[1,*]

and Andrew M. Ellis[3,*]

[1] Institut für Ionenphysik und Angewandte Physik, Universität Innsbruck, Technikerstr. 25, A-6020 Innsbruck, Austria

[2] Institut für Mechatronik, Universität Innsbruck, Technikstr. 13, A-6020 Innsbruck, Austria

[3] Department of Chemistry, University of Leicester, University Road, Leicester, LE1 7RH, UK

Email: paul.scheier@uibk.ac.at; andrew.ellis@le.ac.uk

**Abstract**

Low energy electron attachment to mixed $(H_2)_x/(O_2)_y$ clusters and their deuterated analogues has been investigated for the first time. These experiments were carried out using liquid helium nanodroplets to form the clusters, and the effect of the added electron was then monitored via mass spectrometry. There are some important differences between electron attachment to the pure clusters and to the mixed clusters. A particularly notable feature is the formation of $HO_2^-$ and $H_2O^-$ ions from an electron-induced chemical reaction between the two dopants. The chemistry leading to these anions appears to be driven by electron resonances associated with $H_2$ rather than $O_2$. The electron resonances for $H_2$ can lead to dissociative electron attachment (DEA), just as for the free $H_2$ molecule. However, there is evidence that the resonance in $H_2$ can also lead to rapid electron transfer to $O_2$, which then induces DEA of the $O_2$. This kind of excitation transfer has not, as far as we are aware, been reported previously.

**INTRODUCTION**

Dissociative electron attachment (DEA) has been extensively studied over several decades. It is of fundamental importance in developing an understanding of how electrons interact in collisions with molecules.[1] However, it is also of considerable practical significance given the key role that DEA plays in plasmas. Technological examples of plasma usage include plasma etching,[2] which is extensively employed in semiconductor device fabrication, and the confined plasmas found in experimental nuclear fusion reactors.[3] The plasma state is also highly significant in some astrophysical phenomena, including models of the early universe[4] and in the chemistry of planetary atmospheres.[5] A full understanding of DEA is therefore highly desirable and that necessitates a variety of information including product identification, ion yields as function of electron energy, and reaction cross sections. These have been extensively studied in many gases, including simple diatomics such as hydrogen[6-23] and oxygen.[24-30]

In a recent experiment we reported on the DEA of molecular hydrogen clusters that were isolated in liquid helium nanodroplets.[31] This work led to the first observation of hydrogen cluster anions, $H_n^-$, larger than $H_3^-$. A wide range of cluster anions were detected, all with odd $n$, whose production can be accounted for by DEA of a single $H_2$ molecule in a $(H_2)_m$ van der Waals cluster. A hydrogen atom is ejected from the cluster and the resulting hydride anion, $H^-$, acts as a nucleation centre for the remaining $H_2$ molecules. We also reported clear evidence for the formation of specific structural arrangements within these cluster ions based upon an icosahedral first shell of $H_2$ molecules around the hydride anion.

Here we turn our attention to mixed clusters of $H_2$ (and $D_2$) and $O_2$ in the same environment, namely inside a helium nanodroplet. This has led to observation of electron-induced chemistry between the hydrogen and the oxygen molecules, with strong evidence for the formation of core anions such as $HO_2^-$ and $H_2O^-$. We suggest mechanisms for formation of these ions within the cluster. To assist in this process we have recorded the yields of a wide variety of cluster anions as a function of electron kinetic energy and we see clear resonance processes. We also discuss some unusual electron transfer behaviour which is inferred from the observed resonances.

**EXPERIMENTAL**

Helium nanodroplets were produced by expanding gaseous helium (Messer, purity 99.9999%) at a stagnation pressure of 20 bar through a 5 μm nozzle, which was cooled by a closed-cycle refrigerator (Sumitomo Heavy Industries LTD, model RDK-415D). The nozzle temperature was measured as 9.4 K for the experiments with hydrogen and 9.6 K for the experiments with deuterium, respectively, resulting in both cases in an average droplet size of roughly $5 \times 10^5$ helium atoms.[32] A beam of helium droplets was created by passage of the helium expansion through a conical skimmer with a 0.8 mm diameter aperture. The droplet beam then traversed two differentially pumped pickup chambers: oxygen was introduced in the first chamber and hydrogen (or deuterium) in the second. The doped droplets were subsequently crossed with an electron beam at energies ranging from 0 to 30 eV. The resulting anions were accelerated to 40 eV into the extraction region of a commercial orthogonal reflectron time-of-flight mass spectrometer (Tofwerk AG, model HTOF) with a resolution ($m/\Delta m$) of ~3000. The data were analysed using a custom-built software package, IsotopeFit,[33] which makes it possible to deduce ion abundances from the raw mass spectra. This software takes all possible isotopologues into account and can be used to subtract contributions form background impurities.

**RESULTS AND DISCUSSION**

$H_2$ and $D_2$ have been used in combination with $O_2$ in the experiments carried out for this study. The Supplementary Information[34] shows the results for both $H_2$ and $D_2$ and the results for these two gases are essentially the same. However, for simplicity we focus here on $D_2$. Note also that use of $D_2$ eliminates any potential confusion arising from hydrogen contamination emanating from trace quantities of water in the vacuum system.

The principal anionic products are mixed clusters of the form $D_nO_m^-$. Figures 1 and 2 show how the ion yield varies with $n$ for some selected even and odd values of $m$, respectively. These data were recorded at an electron energy of 12 eV for reasons that will become clear shortly. All of the plots show odd-even intensity oscillations as a function of $n$, particularly for large $n$. Most striking of all is the large ion abundance seen for $n = 1$ in the even $m$ case and for $n = 2$ when $m$ is odd (with the exception of $m = 1$). We shall attempt to explain these observations later.

We next consider ion yields as a function of electron kinetic energy. Figure 3 shows data for the production of $O_2^-$ and $O_4^-$, both with and without added $D_2$ (upper and lower traces, respectively) Resonances are seen, *i.e.* maxima in the ion yield at certain electron energies. Although not shown in this figure, the resonances for odd oxygen anions are similar. The odd oxygen anions are presumably the result of DEA, a process that is well-known for $O_2$, as mentioned earlier. The signal-to-noise ratios for the $O_2^-$ and $O_4^-$ data when $D_2$ is also present in the helium droplet are relatively low but it is nevertheless clear that there are major differences in the ion yield profiles for $O_2^-$ and $O_4^-$ formation in the presence and absence of $D_2$. For pure oxygen clusters, the data from helium droplets essentially agrees with previous gas phase work on oxygen clusters.[29] Thus we see a dominant resonance peaking near 7.5 eV, which is approximately 0.5 eV higher than the gas phase value because of the well-known additional energy needed to inject an electron into a helium droplet.[35] There is also a weaker and broad resonance with a maximum around 14.5 eV. Once again, weak higher energy resonances have been reported for oxygen clusters in the gas phase that match this energy,[29] so in helium droplets the behaviour is similar to the gas phase.

However, once $D_2$ is added the response changes completely. In the $O_2^-$ and $O_4^-$ mass channels (Figure 3(a)) the lowest energy (7.5 eV) resonance is still there but is now overtaken by stronger resonances at higher energy. Specifically, although the signal/noise ratio is not high, there is evidence of two higher energy resonances with maxima near 12 and 13.5 eV. Thus the addition of $D_2$ clearly has a major impact on the ion yields for production of bare oxygen anions. It is interesting to compare the ion yields in Figure 3 with those for the production of $DO_4^-$ in Figure 4. The latter seems to mirror the two high energy resonances seen for $O_4^-$ production in the presence of $D_2$ but the signal-to-noise ratio is much better in this case.

For van der Waals binary clusters, such as $D_2/O_2$ clusters, we would expect to see ion yields that are characteristic of either $D_2$ or $O_2$, with small perturbations as the cluster size increases. This follows from the very weak intermolecular binding, which means that we still have identifiable individual molecules loosely held together. Two ion yield curves are shown in Figure 4, one for formation of $DO_4^-$ and the other for formation of $DO_6^-$. These plots bear no resemblance to those for $O_2^-$ and $O_4^-$ derived from pure oxygen dopant in Figure 3(b). However, they do show some similarity with those from pure deuterium clusters (an example is shown in Figure 3(b) for $D_{25}^-$) in that there is no low energy (sub 10 eV) resonance. In the gas phase $D_2$ shows three resonances below 20 eV which peak at approximately 4, 10 and 14 eV.[15] These resonances lead to dissociative electron attachment in all three cases, i.e.

$$D_2 \; + \; e^- \; \rightarrow \; D \; + \; D^- \qquad (1)$$

The 4 eV resonance disappears when experiments on pure hydrogen and deuterium clusters are performed in helium nanodroplets, something that will be discussed in detail in a future publication. However, the two peaks in Figure 4 match quite closely with the expected resonances for $H_2/D_2$ and their clusters. We are therefore led to the conclusion that the production of $DO_4^-$ is initiated by electron attachment to $D_2$, not $O_2$. Note also that, in terms of ion abundances, we see the same general trends for electron energies of 12 and 14 eV. The data shown in Figures 1 and 2 were taken at an electron energy of 12 eV.

The ion abundances shown in Figure 1 for $D_nO_m^-$ anions with even $m$ all have a very strong signal corresponding to $DO_m^-$. This can be explained by DEA of $D_2$ followed by the reaction of $D^-$ with an $O_2$ molecule, leading to $DO_2^-$. It is well known that the hydride and deuteride anions can react rapidly with $O_2$ via the process:[36]

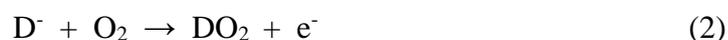

$$D^- + O_2 \rightarrow DO_2 + e^- \qquad (2)$$

Autodetachment normally occurs in the gas phase but in the presence of other molecules and the surrounding helium this process may be inhibited, as has been seen previously in clusters[37] and in the condensed phase.[38] Consequently, we propose that the intense $DO_m^-$ signal is the result of formation of the deuteroperoxy anion, $DO_2^-$, which can be ejected either as the bare ion or can act as a seed for attaching $D_2$ and $O_2$ molecules. The exceptionally high abundance of this ion when compared to the decorated versions, $DO_2^-(D_2)$, suggests that its formation must release considerable energy such that it leads to complete evaporative loss of all $D_2$ molecules in most cases.

Ion abundance oscillations are also seen in Figure 1 as a function of $n$ for fixed $m$. For the case where $m = 2$, $i.e.$ $D_nO_2^-$ ions, the oscillation in abundances can simply be explained by the addition of intact $D_2$ molecules to the core $DO_2^-$ ion. However, there is a surprising change in behaviour when more oxygen is present and this can even be seen in the plot for $D_nO_4^-$. Now the oscillation behaviour is reversed and the more abundant series of anions corresponds to an even number of deuterium atoms. It therefore seems that in the presence of sufficient $O_2$ and $D_2$ the DEA process (reaction (1)) becomes quenched and ion formation favours $D_nO_m^-$ anions composed of intact $D_2$ and $O_2$ molecules.

Turning now to anions with an odd number of oxygen atoms, we see a high probability for the production of $D_2O_m^-$ ions, except when $m = 1$. This is indicative of another chemical reaction taking place, this time leading to the water anion, $D_2O^-$, as the core ion. We again assume that the process begins with resonant electron attachment to $D_2$. Ordinarily this would cause DEA but it is not obvious how the resulting $D^-$ could initiate chemistry leading to formation of $D_2O^-$. We therefore propose that there is competition with the reaction channel described previously. In this new channel the electron attaches to a $D_2$ molecule and then rapidly transfers its charge to $O_2$, a process which may be driven by the higher electron affinity of $O_2$ than for $D_2$. Having accepted this electron, the $O_2$ undergoes DEA to $O^-$, which then undergoes reaction with $D_2$ as shown below:

$$O^- + D_2 + M \rightarrow D_2O^- + M \qquad (3)$$

This process provides a mechanism for formation of $D_2O^-$. Note that the reaction between $O^-$ and $D_2$ has been well studied, is known to be fast, and has essentially no activation energy.[39,40] There are two possible products at low collision energies, $D_2O^-$ and $OD^-$, with the former dominating.[39] Monomeric $D_2O^-$ (and likewise $H_2O^-$ from the corresponding reaction between $O^-$ and $H_2$) is known to be unstable with respect to autodetachment[41] and so produces the neutral molecule, $D_2O$, as the end product. This explains why $D_2O^-$ is not detected in our experiments. However, the anion can clearly be stabilised by the addition of $D_2$ and/or $O_2$ molecules and therefore, for example, we see intense signals for $D_2O_3^-$ and larger $D_2O_m^-$ ions (for odd $m$). This is comparable to the stabilization of $H_2O^-$ by adding a water molecule, which yields an observable $(H_2O)_2^-$ anion in the gas phase.[41] The odd-even oscillations seen in Figure 2 as a function of the number of D atoms can be explained in terms of the DEA of a single $D_2$ molecule, which then combines with intact $D_2$ and $O_2$ molecules, favouring $D_nO_m^-$ ions with odd $n$. If correct, this suggests that there is competition between electron transfer from $D_2$ to $O_2$ (with subsequent DEA) and direct DEA of $D_2$.

What other evidence is there that supports the idea that electron transfer from $D_2$ to $O_2$ can compete with the DEA of $D_2$, and thereby initiate reaction (3)? This brings us back to the ion yield data for pure oxygen anions, such as $O_4^-$, which were discussed earlier. Clearly these oxygen anions can be generated at electron energies which are consistent with resonances for $D_2$ and which are only seen when the $O_2$ is combined with $D_2$ in a helium droplet. We note also that efficient electron transfer between species within clusters has been reported previously. For example, reactions products from the DEA of oxygen clusters show resonance energies characteristic of argon atoms in experiments performed on the mixed clusters $Ar_m(O_2)_n^-$.[42]

**CONCLUSIONS**

The effect of low energy electrons on mixed hydrogen/oxygen clusters has been investigated for the first time. The clusters were formed within liquid helium nanodroplets, which were then irradiated by electrons with controlled kinetic energies. Any anions produced were detected by mass spectrometry. Importantly, the findings from these experiments are not simply a superposition of the expected behaviour for the individual hydrogen and oxygen clusters. Instead we find that the electrons can initiate chemistry and very strong signals from the deuteroperoxy ($DO_2^-$) and deuterated water ($D_2O^-$ ions) are seen. In the latter case the bare $D_2O^-$ ion does not survive because of rapid autodetachment but does survive when combined with at least one $D_2$ or $O_2$ molecule to form a cluster.

The resonance behaviour seen for the hydrogen/oxygen clusters is particularly surprising. The observation of high probabilities for anionic reactions at electron energies near to 12 and 14 eV is interpreted in terms of resonances of $D_2$, since $O_2$ should have no strong resonances in this region. $D_2$ would normally undergo dissociative electron attachment at these energies, and can then react with $O_2$ to yield the $DO_2^-$ ion. However, explaining the formation of $D_2O^-$ is more challenging. Our best explanation at present is that an electron initially undergoes resonant attachment to $D_2$, but that electron is then transferred to $O_2$ before DEA of the former can occur. If correct then there is competition between DEA of the $D_2$ and rapid electron transfer to the $O_2$. When the latter occurs it sets in motion a process which can lead to efficient $D_2O^-$ production.

**Supplementary material**

See supplementary material for a more comprehensive set of anion yield data including for experiments on mixed $H_2/O_2$ clusters.

**ACKNOWLEDGEMENTS**

The authors are grateful for a grant from the Austrian Science Fund, FWF P26635, in support of this work.

**Figure captions**

1. Ion yields as a function of $n$ for $D_nO_m^-$ ions with several different even values of $m$. All traces except that from $D_nO_2^-$ have been vertically shifted for ease of comparison.

2. Ion yields as a function of $n$ for $D_nO_m^-$ ions with several different odd values of $m$. All traces except that from $D_nO^-$ have been vertically shifted for ease of comparison.

3. (a) Dependence of the yields of $O_2^-$ and $O_4^-$ on electron energy for droplets containing a mixture of $D_2$ and $O_2$ in helium nanodroplets. (b) The ion yield curves for $O_2^-$ and $O_4^-$ obtained from pure oxygen clusters in helium droplets and the ion yield curve for $D_{25}^-$ obtained from pure deuterium clusters in helium nanodroplets.

4. Dependence of the yields of $DO_4^-$ and $DO_6^-$ on electron energy.

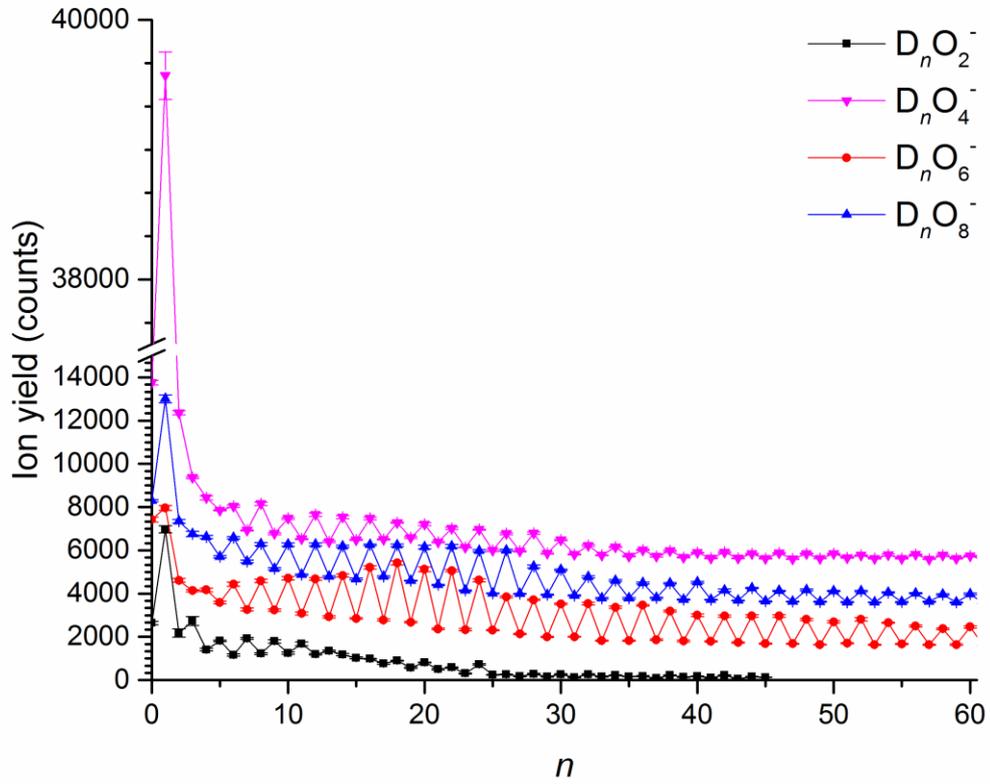

Figure 1

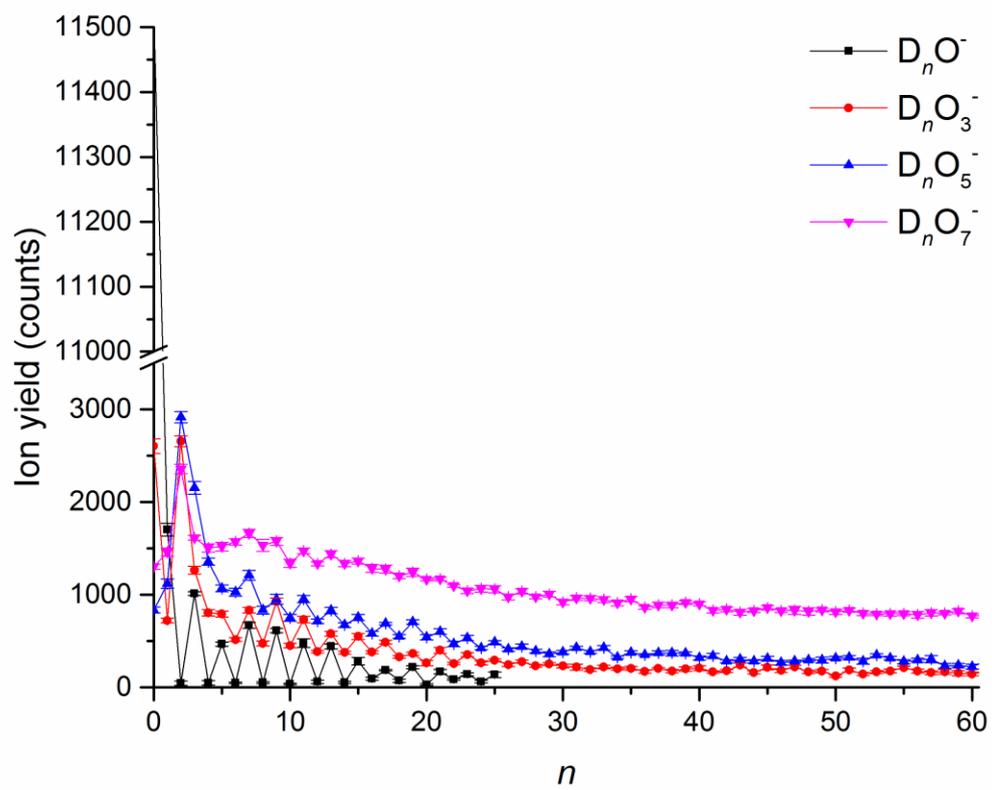

Figure 2

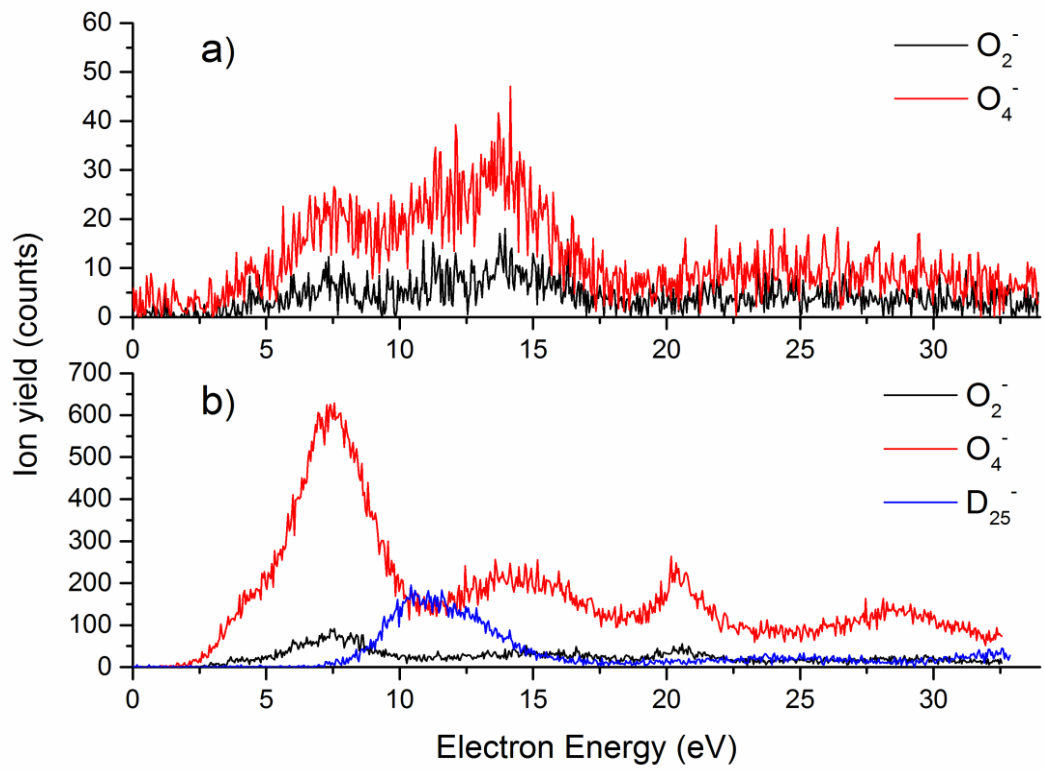

Figure 3

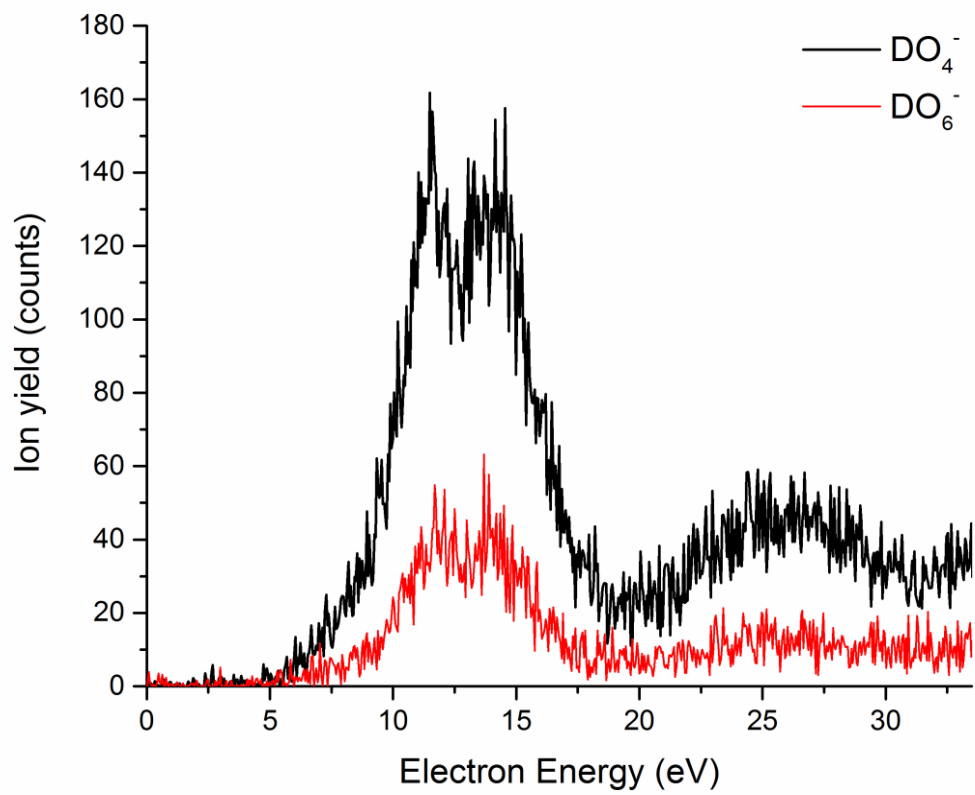

Figure 4